\documentclass[journal]{rmaa}

\usepackage{paralist,ulem}
\usepackage{psfrag,color}

\usepackage[latin1]{inputenc}

\SetYear{2011}
\title{Mass and metal ejection efficiency in disk galaxies driven by
young stellar clusters of nuclear starburst}
\author{Rodr\'iguez-Gonz\'alez,
A.\altaffilmark{1}; Esquivel, A.
\altaffilmark{1}, Raga, A. C.\altaffilmark{1} \& Col\'in, P.\altaffilmark{2}}

\altaffiltext{1}{Instituto de Ciencias Nucleares, Universidad
Nacional Aut\'onoma de M\'exico}

\altaffiltext{2}{Centro de Radioastronom\'ia y Astrof\'isica, 
Universidad Nacional Aut\'onoma de M\'exico.}

\shortauthor{Rodr\'iguez-Gonz\'alez, A. et al.}
\shorttitle{Mass and metal ejection efficiencies.}

\fulladdresses{
\item Esquivel, Alejandro, Raga, Alejandro C. \& Rodr\'iguez-Gonz\'alez, Ary, Instituto de Ciencias Nucleares, Universidad
Nacional Aut\'onoma de M\'exico, Apdo. Postal 70-543,
04510, M\'exico, D.F., M\'exico. (ary@nucleares.unam.mx).
\item Col\'in, Pedro, Centro de Radioastronom\'\i{}a y
  Astrof\'\i{}sica, Universidad Nacional Aut\'onoma de M\'exico,
  Apartado Postal 3--72, 58090 Morelia, Michoac\'an, M\'exico.}

\listofauthors{A. Rodr\'iguez-Gonz\'alez, A. Esquivel, A. C. Raga \& 
P. Col\'in}
\indexauthor{Rodr\'iguez-Gonz\'alez, A.}
\indexauthor{Esquivel, A.}
\indexauthor{Raga, A. C.}
\indexauthor{Col\'in, P.}

\abstract{
We present results from models of galactic winds driven by energy injected
by nuclear starbursts. The total energy of the starburst is provided by 
young central stellar clusters and parts of the galactic
interstellar medium are pushed out as part of the  
galactic wind (in some cases the galactic
wind contains an important part of the metals produced in the new
generation of stars). We have performed adiabatic and radiative 
3D N-Body/Smooth Particle Hydrodynamics simulations of galactic winds 
using the GADGET-2 code. 
The numerical models cover a wide range of starburst 
(from $\sim$10$^2$ to $\sim$10$^7$~M$_\odot$) and galactic gas masses 
(from $\sim$6$\times$10$^6$ to $\sim$10$^{11}$~M$_\odot$).  The concentrated central starburst regions 
are an efficient engine for producing of the mass and metal loss in 
galaxies, and also for driving the metal redistribution 
in the galaxies. }

\resumen{Presentamos las eficiencias de p\'erdida de masa y metales
  obtenidas de modelos num\'ericos de vientos gal\'acticos empujados
  por la energ\'ia depositada en brotes de formaci\'on estelar
  nucleares. Los brotes de formaci\'on estelar contienen
  c\'umulos estelares j\'ovenes los cuales inyectan la energ\'ia
  suficiente para empujar parte del medio interestelar fuera de las
  galaxias. En algunos casos los vientos gal\'acticos contienen una
  importante parte de los metales producidos por las nuevas
  generaciones estelares. Para estudiar las eficiencias de p\'erdida
  de masa y metales hemos desarrollado simulaciones num\'ercias 3D
  ``N-Cuerpos/Smooth Particle Hydrodynamics'' de vientos gal\'acticos
  para los casos: adiab\'aticos y con p\'erdidas radiativas.  Los modelos
  num\'ericos cubren una amplio intervalo de masas en los brotes de
  formaci\'on estelar (de $\sim$10$^2$ a $\sim$10$^7$ M$_\odot$) y de
  masas  en las galaxias anfitrionas (de $\sim$6$\times$10$^6$ a  
$\sim$10$^{11}$ M$_\odot$).  Las regiones de formaci\'on estelar concentradas en el centro
  del potencial, son una maquinaria importante para la p\'erdida y
  redistribuci\'on  de masa y metales en este tipo de galaxias.}

\addkeyword{stars: winds}
\addkeyword{galaxies: star clusters}
\addkeyword{galaxies: starburst}
\addkeyword{ISM: general}

\begin{document}
\maketitle

\section{Introduction}

A great amount of evidence both, observational and theoretical, 
indicates that Galactic Winds (GWs) are a necessary
and very important ingredient of the evolution of galaxies
and the inter-galactic medium. Their
presence can explain, among other things, the very low metallicities found 
in dwarf galaxies or the abundance of metals observed in the
intra-cluster medium. 
Such outflows have been observed in galaxies of a variety of masses
and at different redshifts (e.g., Martin 1999, 
Pettini et al. 2001, Veilleux et al. 2005). At the same time,
different analytical and numerical  models have been
put forward to explain these outflows (e.g., Chevalier \& Clegg 1985,  
Tomisaka \& Ikeuchi  1988,
Heckman et al. 1990, De Young \& Heckman  1994, 
MacLow \& Ferrara 1999, Strickland \&
Stevens 2000, Cant\'o et al. 2000, Tenorio-Tagle et al. 2003,
Fujita 2004, Cooper et al. 2008, Fujita et al. 2009, 
Rodr\'iguez-Gonz\'alez et al. 2009).

Long ago, Lynds \& Sandage (1963) found a large outflow in M\,82 and gave the
first  definition of a starburst (SB). They  
proposed that material was ejected from the nuclear regions as a consequence 
of a series of SN explosions. Since then, galaxies with concentrated 
central starbursts (M\,82, NGC\,253, NGC\,1569, to name a few) have been used 
to study the GW phenomenon.

Tenorio-Tagle et al. (1998) showed the importance of the galactic potential 
component due to the cold dark matter halo in realistic models of GW, and
concluded that the fate of the stellar wind and SNs material depends
on the galactic properties and on the total energy injected by the SB. 
They showed that dwarf galaxies with an interstellar medium (ISM)
with masses of the order of 10$^9$ M$_\odot$, retain their metals
unless they undergo  an extreme burst of star formation, much larger
than those presently observed. On the other hand,
MacLow \& Ferrara (1999) explored both, analytically and numerically,
the effects of stellar winds and SN explosions, in SBs with mechanical
luminosities in the $\sim$10$^{37}$ - 10$^{39}$ erg s$^{-1}$ range, on
the ISM of dwarf galaxies with gas masses  in the  $\sim10^6$ to
$10^9$ M$_\odot$ range. They distinguished between two
  possibilities of  gas ejection: a) A ``blowout'', which require material 
be accelerated above the escape speed at a distance of about three times the 
scale height. b) A ``blow-away'', in which virtually all the ISM is 
accelerated and lost to the galaxy.
The blowout would happen preferably for flatter galaxies,
rounder galaxies are more likely to be completely disrupted
(blow-away) if enough energy is injected.  
Their conclusion was that dwarf galaxies in that mass range undergo
blowout for moderate-to-high luminosity starbursts values ($L \sim 10^{38}$-
$10^{39}$~erg s$^{-1}$), whereas for galaxies with gas masses 
$M_g > 10^{8}$~M$_\odot$ and mechanical luminosity 
$L \sim 10^{37}$~erg s$^{-1}$ the blowout was inhibited. They also found 
that in the lowest mass objects, blow-away occurs virtually independently 
of $L$.

Many authors have shown that in the nearest starburst galaxies GWs could 
form through the collective effect of many individual stellar cluster  
winds, which in turn  are formed by the collective effect of many individual 
stellar winds. Perhaps, the best studied starburst galaxy is M\,82, which 
has an extended, biconical filamentary structure in the optical  
(Shopbell \& Bland-Hawthorn 1998; Ohyama et al. 2002). In M\,82, the
super stellar clusters (SSCs) H${\alpha}$ luminosities are in the  
(0.01-23)$\times 10^{38}~\rm{erg~s^{-1}}$  range (Melo et al 2005). These 
authors compiled a catalog of 197 SSCs  in M\,82 and 48 SSCs in NGC\,253 
(Melo 2005), with masses in the $10^4< \rm{M/M}_{\odot}<10^6$ range (more 
recently, additional young star clusters in M\,82 have been reported by Smith
et al. 2007 and Westmoquette et al. 2007). Other examples are NGC\,1569 
(Anders et al. 2004; Westmoquette, Smith \& Gallagher III 2008) and  M\,83 
(Harris et al. 2001), with SSCs densities and masses similar to
those of M\,82. Such an exceptional density of massive clusters (i.e
$\sim 620~\rm{kpc}^{-2}$, see also Anders et al 2004; Melo et al. 2005;
Melo 2005) is the best scenario to study powerful  
GWs driven by SSC winds originated in nuclear starburst regions. 

Yet, another interesting effect is when the material ejected from
the SSCs winds does not abandon the galaxy. 
Tenorio-Tagle 1996 explored the effect of a non uniform interstellar medium 
(density and temperature) in the cooling efficiencies within the giant 
cavities. In high density regions the cooling would be more effective and would
produce a series of metal-rich cloudlets. Such cloudlets would fall
back to the disk and produce local enhancements of metallicity. However, 
Recchi et al. (2001; 2008) presented arguments against this scenario, pointing out 
that thermal conduction and the hydrodynamical drag are likely to disrupt the 
condensations before they reach the disk, and thus the metals would be spread 
more or less uniformly  throughout the galaxy.

In the present paper we study the effects of young stellar 
clusters in the nuclear starburst (SB) region on the ISM of dwarf 
disk-galaxies. In particular, we calculate the amount of enhanced metallicity
material that will end up in the intergalactic medium (IGM) and/or in
the outer regions of the galactic disk.
The paper is organized as follows: in \S 2 we describe the
ingredients of the galactic models that we use, in \S 3 we discuss the
possible fate of the  mass and metals in such galaxies. In \S 4 we
describe a series of 3D simulations of the interaction of SSCs
with the galactic ISM. The results are presented in \S 5,  and our
conclusions in \S6.

\section{Galactic model ingredients}

Each galaxy in our study consists of a dark matter halo, and a 
rotationally  supported disk of gas and stars. The models are 
constructed following the approach described in 
Springel et al. (2005, see also Hernquist 1993 and Springel 2000).
A near-equilibrium galaxy model is constructed, with a
dark matter halo that follows a Hernquist (1990) profile, and a disk with 
exponential surface gas density. Then, using an iterative procedure,
the vertical gas profile is  determined self-consistently for a
particular effective equation of state.

\subsection{The halos}

We modeled the dark matter mass distribution with the Hernquist
profile (Hernquist 1990),
\begin{equation}
\label{rhoh}
\rho_{h}(r)=\frac{M_{dm}}{2\pi}\frac{a}{r(r+a)^3},
\end{equation}
where $r$ is the radial distance, and $M_{dm}$ is the dark matter mass.
Two remarks can be made about the shape of this profile: i) in 
the inner parts of the galaxy it agrees with the profile given by 
the Navarro, Frenk \& White (1996, henceforth NFW) fitting formula. 
ii) Its faster than $r^{-3}$ decline makes the total mass to converge, 
allowing the construction of isolated halos without the
need of an ad-hoc truncation. The $a$ parameter is related to 
the scale radius $r_s$ of the NFW profile by:
\begin{equation}
\label{a}
a=r_s\sqrt{2[ln(1+c)-c/(1+c)]}\,,
\end{equation}
where $c=r_{200}/r_s$ is the concentration parameter, $r_{200}$ is the radius 
at which the enclosed dark matter mean density is 200 times the critical
value (where the critical density is $\rho_{crit}=3H^2/8 \pi G$).
We have set $c = 9$.

\subsection{The disk}

We modeled gas and star disk components, with an exponential
surface density profile (in the radial direction of the
galactic disk, $R$) the scale of length $R_0$,
\begin{equation}
\label{sigmag}
\Sigma_{g}(R)=\frac{M_{g}}{2\pi R_0^2}\exp(-R/R_0),
\end{equation}
\begin{equation}
\label{sigmas}
\Sigma_{*}(R)=\frac{M_{*}}{2\pi R_0^2}\exp(-R/R_0),
\end{equation}
so that the total mass of the disk is $M_d=(M_g+M_*)=m_d M_{tot}$, 
where $m_d$ is a dimensionless parameter (fixed in this work to
$m_d=0.041$), and $M_{tot}$ is the total mass of the galaxy, incuding
the dark matter halo (i.e. $M_{tot}=M_d+M_h$, where $M_h$ is the mass
of the halo).
The vertical mass distribution of the stars in the disk is specified
by the profile of an isothermal slab with a constant scale 
height $H$. The 3D stellar density in the disk is hence given by,
\begin{equation}
\label{denstar}
\rho_* (R,z)=\frac{M_*}{4 \pi\,H\,R_0^2} {\rm sech}^2\left(\frac{z}{2H}\right)
\exp\left(-\frac{R}{R_0}\right).
\end{equation}
In all models, the $H/R_0$ was fixed to $0.2$ whereas
the relative content of gas in the disk, the rest of the disk
mass is in stars, was set to $0.35$. On the other hand,
the disk spin fraction (fraction of the total angular 
momentum in the disk), was set, as it is customary, to the value of $m_d$.
These values are typical of low-mass galaxies (e.g., Springel et al. 2005).

\subsection{Vertical structure of the gas disk}

The vertical structure of the disk at the beginning of the simulation
is computed self-consistently with the halo and stellar disk
components as follows. At $t=0$ the gaseous disk has uniform
temperature (fixed to $1000~\mathrm{K}$).
Consider the vertical structure of the gas disk is in hydrostatic equilibrium:
\begin{equation}
\label{hidros}
\frac{\partial \rho_g}{\partial z}=-\frac{\rho^2_g}{\gamma P}
\frac{\partial \Phi_T}{\partial z}
\end{equation}
where, $\Phi_T=\Phi_h+\Phi_d$ is the total gravitational potential 
($\Phi_h$ and $\Phi_d$ are the halo and disk gravitational potentials,
respectively).
For a given $\Phi_T$, the solution of this equation is constrained
by the condition
\begin{equation}
\label{sigmag2}
\Sigma_g(R)=\int \rho_g(R,z)dz\,,
\end{equation}
where $\Sigma_g(R)$ is the gaseous surface mass density (see also
eq.~\ref{sigmag}). 
One can obtain the vertical distribution by integrating
eq. (\ref{hidros}), with a central density value at the midplane
($z=0$) and an effective equation of state of the
  form $P=P(\rho)$, see Springel et al. 2005.)
The density starting value is guessed, then adjusted in an
iterative scheme until the desired surface density (eq. \ref{sigmag})
is recovered. This process is repeated for different radii to obtain
an axisymmetric  gas density distribution.

\subsubsection{Cooling}
For the non-adiabatic cases, we adopt the cooling functions 
computed in collisional ionization equilibrium (CIE) by Sutherland 
\& Dopita (1993). With the energy loss per unit mass of the gas by,
\begin{equation}
\label{cool}
\left(\frac{du}{dt}\right)_{cool}=-\frac{\Lambda_{net}(\rho,T)}{\rho}.
\end{equation}

We must note that the cooling function we used in our models assumes
solar abundances. However, the kind of galaxies treated in this paper 
should be less metallic and thus the cooling should be lower. For the
less metallic galaxies the results would shift from our radiative
models to the adiabatic ones.

\section{The fate of the gas injected by young stellar clusters}\label{sec:fate}

Depending on the total amount of kinetic energy injected by
the SSCs two outcomes are possible: if the kinetic energy
is larger than the gravitational potential of the galaxy, all 
material will escape and integrate to the IGM; otherwise, the
gas will return to the disk. 
If the material returns and reincorporates to the
galaxy two things can happen: it can fall back close to the
galactic center or far from it. Considering that GWs are confined 
into a cone, material that does not reach a high
altitude will fall close to the galactic center, while gas that
reaches a high altitude can fall back at large galactocentric radii.

\subsection{Mass loss measures}

In order to quantify the mass loss in numerical models 
one has to determine which material will be considered lost to the
IGM. A natural estimation can be derived by comparing the velocity of
a parcel of gas and the escape velocity at its current position. This
has been done for instance in the previous works of MacLow \& Ferrara
(1999) and D'Ercole \& Brighenti (1999). This works, however, used
Eulerian codes in which material that leaves the computational domain
can not be further analyzed, and therefore was automatically
considered unbound. In reality, some
of this material would return to the galaxy. Our models, that
use the SPH (Smooth Particle Hydrodynamics) integration scheme, 
are not limited by the extent of the
computational domain, they thus were able to 
trace all the material for the duration of the simulation. We will
consider the following more stringent criterion:
only material that has left a cylindrical region of R$_{max}$=20~kpc
and $z_{max}$=$|20|$~kpc (see also  D'Ercole \& Brighenti 1999) {\it
  and} has a velocity larger than the escape velocity at its current
position will be considered effectively lost.

Mass loss in our models are thus carried out by first 
computing the total mass of unbound (ejected from the galaxy) gas $M_{ej}$. To
do this, we consider the total galaxy  gravitational potential 
given by,
\begin{equation}
\label{phit}
\Phi_T(r)=\Phi_h(r)+\Phi_d(r),
\end{equation}  
where $r$ is the distance to the center, and $\Phi_h$ and $\Phi_d$
are the dark matter halo and disk components, respectively. 
Gas particles outside the cilinder with velocity larger than the escape velocity
\begin{equation}
  v_{esc}=\sqrt{-2\Phi_T(r)},
\label{vesc}
\end{equation}
are considered to be unbound.
We define a `total mass ejection efficiency' as the ratio of the
mass ejected (unbound) to the total gas mass,
\begin{equation}
\label{efmass}
\xi_m \equiv \frac{M_{ej}}{M_g}.
\end{equation}
The mass ejection efficiency has two limiting  cases: $\xi_m \to 0$
when all the galactic ISM is bound, and $\xi_m \to 1$ when all the
galactic gas is expelled from the galaxy.

One can also define a ``metal ejection efficiency'' as 
\begin{equation}
\label{ej}
\xi_z\equiv\frac{M_{c,ej}}{M_c},
\end{equation}
where $M_{c}$ is the total mass injected by the SSCs, and $M_{c,ej}$ the mass,
also originated in the SSCs that is unbound. The mass injected by a
star cluster contains  
the metals produced in the lifetime of the massive stars, $t\sim 40$ Myr. 
For that time-span, using the metal yields calculated by Meynet \& Maeder 
(2002), Tenorio-Tagle et al. (2005) showed that the metallicity of the combined
stellar winds remains at a value of $\sim 0.5\, Z_\odot$ from the start
of the SB, up to a $t=3$~Myr evolutionary time, and then rapidly
grow to  $\sim 15\, Z_\odot$ at $t\approx 7$~Myr, to decrease gradually
to  $\sim 3\, Z_\odot$ at $t\approx 20$~Myr.

The metal ejection efficiency has two limiting cases: $\xi_z \to 0$, when
all the metallic gas injected by the nuclear starburst is kept in the 
galaxy disk, and $\xi_z \to 1$ case, when all the metallic gas 
processed in the burst is expelled from the galaxy. Galaxies with 
$\xi_z \to 1$ would be metal poor, while the IGM around them should be
metal rich.
On the other hand, metallic gas injected from the SBs region 
with $v_g < v_{esc}$ is going to return to the galactic plane, and the
ISM of such galaxies will be contaminated with the metals produced in
the nuclear  starburst.

\subsection{Disk contamination measures}

As mentioned above, bound gas ejected by the SSCs will return to the
galactic disk, either close to the center if it never reached a high
altitude, or spread to large galactocentric radii if it did.
This mechanism is of course a continuum, and ``high'' or ``low'' altitude
are just relative terms. To quantify the contamination of the disk by
material originated in the SSCs we will consider as `high altitude' a
distance of at least 3 times the scale height of the disk, $H$.

One can define a new parameter, the fraction of bound
  gas (that is inside a cylinder of radius R$_{max}$
  and with velocity less than the escape velocity) that
  reaches at least an altitude of $3H$:
\begin{equation} 
\label{reins}
\chi_b\equiv\frac{M_{ej,3H}}{M_g}\,,
\end{equation}
where $M_{ej,3H}$ is the total bound mass (including the ISM and the material 
from the SSCs) that reaches an altitude of $\geq 3H$.
This mass will be reinserted to the galactic disk at different
radii. However, since it includes material both from the SSCs (with
high metallicity) and from the ISM around the SSCs (lower metallicity)
it is not clear how much the metallicity of the disk will be enhanced
by this mixture. For that reason we define the fraction of
metallic bound gas that leaves the galactic disk and reaches an altitude 
between $3H$ and z$_{max}$ as,
\begin{equation} 
\label{reinsz}
\chi_{b,z}\equiv\frac{M_{c,3H}}{M_c},
\end{equation}
where $M_{c,3H}$ is the bound mass injected by the stellar cluster
that reaches an altitude $3H \leq z \leq z_{max}$, and $M_{c}$ the
total mass injected by the SSC wind.
This gives a better measure
of the amount of high metallicity material available to spread
throughout the disk. Nevertheless, a detailed study of how it will
be distributed in the disk is necessary for a complete disk contamination
model, which is beyond the scope of this paper.

\begin{figure*} 
\includegraphics[width=14cm]{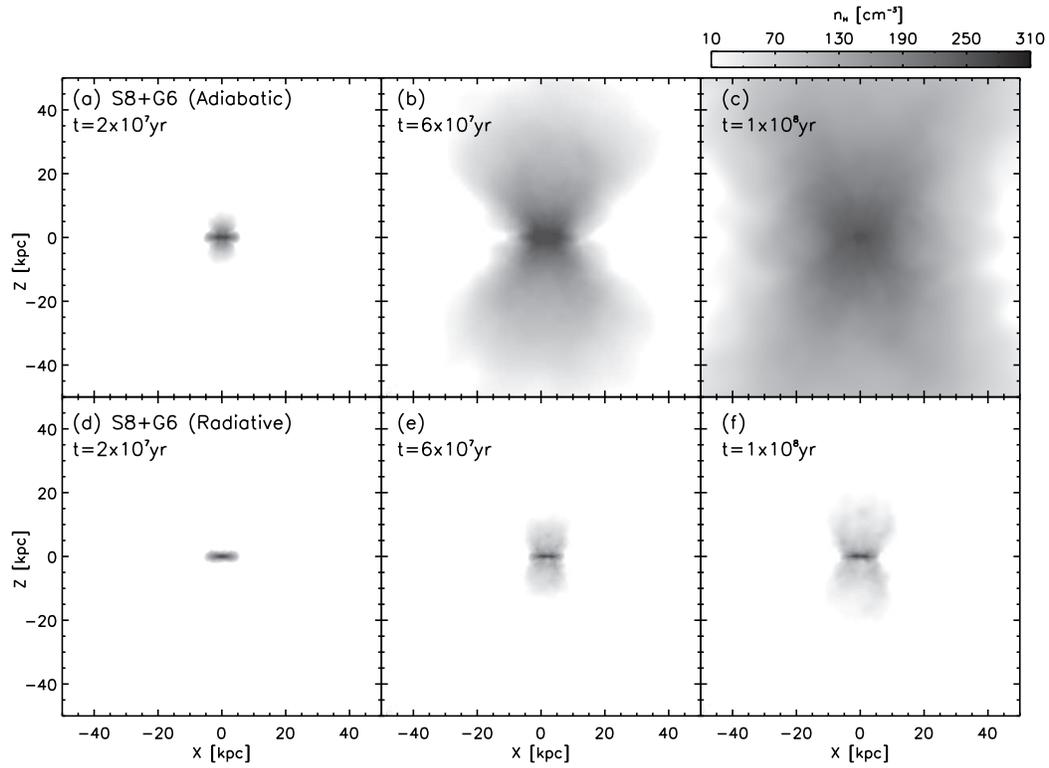}
\caption{Time sequence of cuts of  number density in $xz$-midplane
($y=0$) for the model S8+G6 (see Tables~1\&2). The integration
times in years are given in the upper left of the frames.  The top panels 
(a, b and c) are the adiabatic models, and  the bottom panels are the radiative 
ones.}
\label{s8g6}
\end{figure*}
\section{Numerical simulations}

We have performed adiabatic and radiative 3D N-Body/SPH simulations
that model the effect of a
central compact starburst, in which mechanical energy is injected by
stellar cluster winds,  using GADGET-2\footnote{GADGET-2 is a code 
for cosmological N-Body/SPH-simulation on serial workstations or 
massively parallel computers with distributed memory (Springel et 
al. 2001 \& Springel 2005)}. All simulations 
were done with 3 types of particles: disk, halo, and gas particles,
and,  unless otherwise stated, each consist of 30000, 40000, and 
30000 particles, respectively. Two separated simulations were run for
each model, one adiabatic and a second one including radiative losses.

Most simulations (low-resolution simulations) use a timestep 
$\sim 1-3\times 10^5$~yr and a Plummer softening of 0.25~kpc, 
0.5~kpc, and 0.25~kpc, for gas, halo, and disk particles,
respectively. We ran a grid of models using different galaxies masses 
and starburst energies. The parameters of the models are described below.

\subsection{The host galaxies}

We constructed ten isolated galaxies as described in Springel (2000) and
in \S 2 of this paper. The galaxies
are named from G1 to G10, they have ISM masses in the range of  $\sim 
5 \times 10^6$ - $10^{10}$~M$_\odot$, similar to those used by
MacLow \& Ferrara (1999), the total masses are in the $\sim 4
\times 10^{8}$ -  $1 \times 10^{12}$~M$_\odot$ range. 
The disk masses of the galaxies are given by $M_d=0.041M_{tot}$, and the
total mass of the gas is of 35 \% of the galactic
visible mass in all of  the models.

Table~1 shows the mass of the gas and the halo, the circular 
velocity at r$_{200}$ and $R_0$  for all the galactic models.
\begin{table}
\centering
\caption{The galactic models.}
\begin{tabular}{lcccccc}
\hline
Galaxy & $M_{g}$        &    $M_h$           & $v_c(r_{200})$ & $H$ & $N^a_{ep}$ & $N^r_{ep}$ \\
model  & M$_{\odot}$    &    M$_{\odot}$     & km/s  &  kpc   &       &\\
\hline
G1 & $6.2 \times 10^6$ & $4.3 \times 10^8$ & 12      & 0.044 & 335    & 478\\
G2 & $3.0 \times 10^7$ & $2.1 \times 10^9$ & 20.8    & 0.071 & 530    & 766\\
G3 & $7.0 \times 10^7$ & $4.9 \times 10^9$ & 27.7    & 0.094 & 473    & 573\\
G4 & $1.4 \times 10^8$ & $9.4 \times 10^9$ & 34.3    & 0.118 & 458    & 646\\
G5 & $2.1 \times 10^8$ & $1.4 \times 10^{10}$&39.4   & 0.135 & 453    & 681\\
G6 & $3.0 \times 10^8$ & $2.1 \times 10^{10}$&44.5   & 0.153 & 429    & 666\\
G7 & $4.7 \times 10^8$ & $3.2 \times 10^{10}$&51.9   & 0.178 & 379    & 655\\
G8 & $7.0 \times 10^8$ & $4.8 \times 10^{10}$&59.2   & 0.203 & 317    & 585\\
G9 & $1.1 \times 10^9$ & $7.3 \times 10^{10}$&68     & 0.233 & 255    & 434\\
G10& $1.5 \times 10^{10}$& $1.1 \times 10^{11}$&76.8 & 0.263 & 191    & 317\\
\hline \hline
\end{tabular}
\end{table}

\subsection{The starburst regions}

In order to reproduce the observed properties of starburst galaxies with 
young stellar clusters, including  M\,82, NGC\,253 and 
NGC\,1569, we use a wide range of SBs masses. We should note that the 
number of these energetic stellar asociations in the host galaxies 
were detected at optical wavelengths (e.g. for M\,82  see Melo et al. 2005; 
for  NGC\,1569 see Westmoquette et al. 2008). Since there is high a level of 
obscuration in the nuclear regions of M\,82, NGC\,1569, and NGC\,253, the
 real number of clusters would be larger. Melo et al. (2005) and Smith
et al. (2007) have shown that in the nuclear SBs region of M\,82 the
SSCs are confined to the central $\sim$ 200 pc of the galaxies.

We estimated the mass of the starburst using Starburst 99 models 
(Leitherer \& Heckman  1995 and Leitherer et al. 
1999\footnote{http://www.stsci.edu/science/starburst99/}) 
with the appropriate mechanical energy and a Salpeter initial 
mass function (IMF) that includes stars between 1 and 100 M$_\odot$. 
In order to obtain a better estimation of the total starburst masses,
we have calculated the mass of the stars between 0.1 and 1 M$_\odot$
using the Kroupa et al. (1993) IMF. 
We assumed that all of the energy produced in the SB lifetime
is injected instantaneously, at a given time $t_c$,
inside a spherical region of radius $R_c$ at the center of the disk
galaxy as thermal energy. At this time (corresponding to
$\sim 10$ orbital periods of the galactic disk at
a radius of 150~pc), the gas in the central regions of the
galaxy has already developed a clumpy structure. Actually, in the
present version of the code we cannot consider the case of
an injection that extends over a finite time period.

The total injected energy is obtained from
\begin{equation}
\label{totalenergy}
E_{c}= \int^{tf}_0 L_m(t) dt,
\end{equation}
where $t_f$ is the typical SBs lifetime,  and  $L_m(t)$
the mechanical luminosity as a function of time. The most 
important contribution to the mechanical luminosity happens at
$\sim 40$ Myr (see also Leitherer et al. 1999) and after that time the
mechanical luminosity decreases drastically, due to the fact that 
the most massive stars have died by then. Table~2 shows the total
energy and the respective mass of all the starburst models (S1-S11). 

\begin{table}
\centering
\caption{The starburst models.}
\begin{tabular}{lccccc}
\hline
Starburst    &  $E_{c}$ & $M_{SB}$  \\
Model        &  erg       & M$_\odot$  \\
\hline
S1  &$1 \times 10^{52}$ & $\sim 5.5 \times 10^2$\\
S2  &$5 \times 10^{52}$ & $\sim 2.75\times 10^3$\\
S3  &$1 \times 10^{53}$ & $\sim 5.5\times 10^3$\\
S4  &$5 \times 10^{53}$ & $\sim 2.75\times 10^4$\\
S5  &$1 \times 10^{54}$ & $\sim 5.5\times 10^4$\\
S6  &$5 \times 10^{54}$ & $\sim 2.75\times 10^5$\\
S7  &$1 \times 10^{55}$ & $\sim 5.5\times 10^5$\\
S8  &$5 \times 10^{55}$ & $\sim 2.75\times 10^6$\\
S9  &$1 \times 10^{56}$ & $\sim 5.5\times 10^6$\\
S10 &$5 \times 10^{56}$ & $\sim 2.75\times 10^7$\\
S11 &$1 \times 10^{57}$ & $\sim 5.5\times 10^7$ \\
\hline \hline
\end{tabular}
\end{table}

\subsection{The grid of models}

We computed a total of 220 numerical simulations of GWs
produced by the nuclear starburst events, with the
different total  masses of the galaxies G1-G10 (see Table 1), combined with the
different starburst  energies S1-S11 (see Table 2), all adiabatic as well 
 radiative (110 models each).
The total energy injected by the starburst of the models
S1-S11 was distributed in the gas particles inside
a sphere of $R_c=150$ pc at the center of each of the galactic models
G1-G10.

We choose this size for the region in which the energy
of the starburst is injected, because it is representative of the
sizes of observed starbursts. For example, in M\,82 the  starburst
region has a $R_c \sim 175$ pc radius (Moorwood 1996), and in NGC\,1569A
the starburst hole has a $R_c\sim 100$ pc radius (Fujita et al. 2009).
An exploration of the effects on our results of varying $R_c$ is
shown in Section 6.
 
The number of energetic particles ($N^a_{ep}$ and $N^r_{ep}$, for the
adiabatic and radiative cases, respectively) is related to the 
total number of particles of gas in the initial conditions.
The energy per particle in the starburst region is $E_i=E_c/N_{ep}$, 
the number  of energetic points used in the galactic models,
is shown in Table~1. All the cluster particles (the energetic 
particles) are tagged in order to estimate the fate of the metals
produced during the starburst lifetime.

\begin{figure}
\centering
\includegraphics[width=\columnwidth]{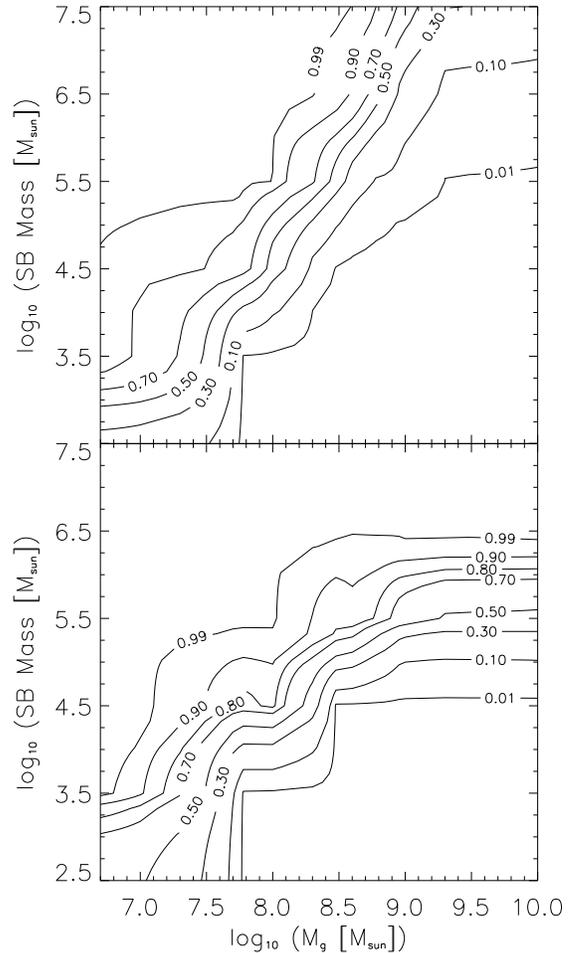}
\caption[]{Mass ($\xi_m$, top) and metal ejection efficiency ($\xi_z$, 
bottom) countours as a function of total mass of the gas and total starburst 
masses, for adiabatic models (upper and lower panel, respectively).}
\label{fig:efa}
\end{figure}
\section{Results}

We have let the galaxy models evolve on their own for
$t_c=1.5\times10^7$ yr. At the time $t_c$, the total energy  of the SB 
is injected in the center of the galaxy. MacLow \& Ferrara (1999)
injected the energy from the cluster in a continuous way over a period
of 40~Myr. 
Given the fact that the flow induced by the cluster takes $\sim
0.5$~Gyr to travel a distance of $\sim 10$~kpc (i. e., the size of the
galaxy), the differences that might result from releasing the cluster
energy over $\sim 40$~Myr=0.04~Gyr (rather than instantaneously)
cannot be very important. 

The ejection efficiencies were computed at $t=1$~Gyr in all our models.
The values of the efficiencies reach a constant value of integration
times of $\sim$0.2-0.5~Gyr and $\sim$0.4-0.7~Gyr for the
adiabatic and radiative models (respectively).

\subsection{Mass and metal ejection efficiencies}\label{sec:ejec}
\begin{figure}
\centering
\includegraphics[width=\columnwidth]{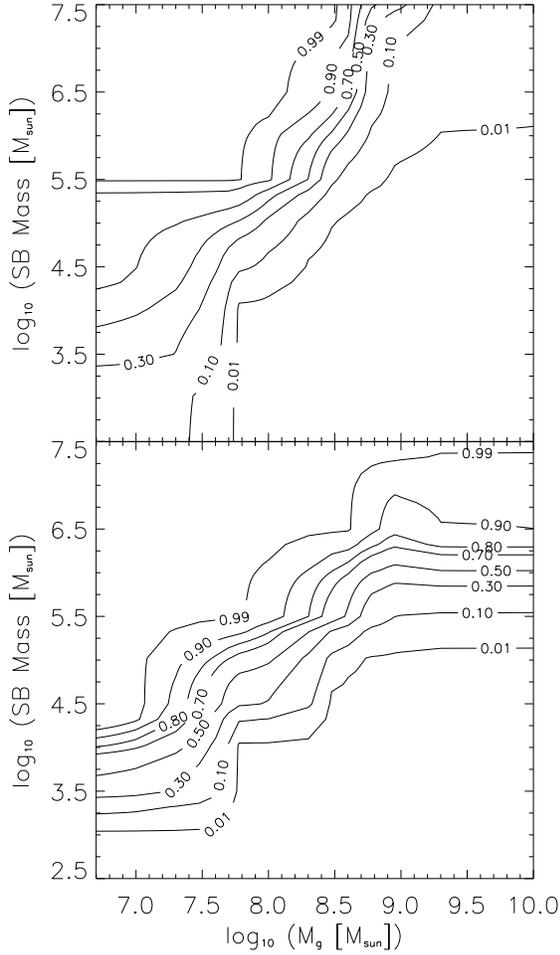}
\caption[]{Same as Figure~2, but for the radiative models.}
\label{fig:efr}
\end{figure}
Total mass and metal ejection efficiencies have been calculated using the 
definitions given in section~\ref{sec:fate}. In particular, 
the kinetic and gravitational potential energies as a function of the
radial distance $r_i$ (with respect to the center of the potential)
were computed for each gas particle.
A particle $i$ would escape the galactic gravitational potential if
$v_i \geq  v_{esc}=\sqrt{-2 \Phi_i}$, where $v_i$  is the
velocity  and $\Phi_i$ the total
potential at the position of the particle.

We also traced the high metallicity gas particles (injected by the
SBs) and  low metallicity gas (galactic ISM gas).
The high metallicity gas was  tagged with a negative passive scalar,
the ISM with a positive one.
Therefore, $\xi_z$  was calculated using the gas particles  with a negative
passive scalar, while $\xi_m$  was calculated with all the gas
particles (regardless of the sign of the passive scalar).

Figure~\ref{s8g6} shows the temporal evolution of the ejected gas 
from a central starburst with properties that
correspond to starburst  model S8 at the center of  galaxy model G6 (see
Tables~1 \& 2).
The  three timesteps shown are
$2\times10^7$, $6\times10^7$  and $10^8$ yr, for the 
adiabatic case --panels (a), (b) and (c), respectively--, and the
radiative one --(d), (e) and (f), respectively--.  
We can see the ejection of gas from  
the plane of the dwarf galaxy with a moderate ISM mass ($3\times10^8$
M$_\odot$), produced by a central starburst  that ejects a mass of
$1.5\times10^6$ M$_\odot$. Such a SB corresponds to a region
with $\sim 9000$ massive stars, injecting a total energy  
during its lifetime of around $5\times 10^{55}~\mathrm{erg}$ 
(see the starburst models of Leitherer \& Heckman 1995 and  Leitherer et 
al. 1999).
From the Figure we see right away that the amount of gravitational 
unbound gas of the galaxy is larger in the adiabatic than in the
radiative one.  However, in both cases an important part of the disk
gas is pushed above the galactic plane, and in the radiative case a
significant part of this gas is returned to the galactic disk.

For this model (S8+G6) the totality of the galaxy  gas is pushed above $3H$, 
however, only $43\%$ acquires enough energy to leave the galactic potential 
and integrate with the IGM, while the remaining $57\%$ would eventually 
return to the disk of the galaxy. However, this ISM material is gravitationally
bound and is going to return to  the galactic plane.
For the radiative case, the percentage of the gas that is gravitationally
unbound is of 14\%, while 4.7\% of the ejected gas gets above 3$H$
and returns to the galactic disk. On the other hand, in the adiabatic case 
97\%\ of the metallic gas ejected by the nuclear starburst escapes  
from the galactic gravitational potential. 
In other words, a galaxy with those parameters would be
contaminated only with 5\% of the metals produced in the massive SB. For the 
radiative S8+G6 model, 79\%\ of the metals escape from the galaxy.

The mass ejection efficiencies are presented in Figures~\ref{fig:efa} 
and \ref{fig:efr}, for the adiabatic and the radiative
model grids, respectively. In these figures,  the abscissa sweeps 
the galactic models (G1-G10), and the ordinate sweeps the SB models 
(S1-S11).

Figure~\ref{fig:efa} shows isocontour maps of the mass ejection efficiency
$\xi_m$ (top panel) and the metal ejection efficiency
$\xi_z$ (bottom panel) for the adiabatic models as a function of the SB mass 
and the total gas mass of the galaxy. For the mass ejection 
efficiency (top panel) we plotted isocontours for $\xi_m=$ 0.01, 0.10,
0.3, 0.5, 0.7, 0.9 and 0.99. In this figure, we can see that for the more
massive galaxies (in which the gas mass is larger than
10$^9$M$_\odot$) the fraction of unbound gas is less than $\sim 
0.25$, even for a rather large nuclear starburst. However, galaxies 
with an ISM of mass lower than $10^8$M$_\odot$ will lose virtually
all of their gas if they undergo a nuclear starburst that injects more
than 10$^5$M$_\odot$.

From the $\xi_z=$ 0.01, 0.10, 0.3, 0.5, 0.7, 0.8, 0.9 and 0.99 isocontours 
in the bottom panel one can see that most of the gas ejected from the cluster
escapes the galaxy if the starburst mass is $>10^{6.5}$M$_\odot$.
Also, most of the ejected gas escapes if the galaxy has
a total gas mass $<10^7$M$_\odot$. On the other hand, for starburst
masses $<10^{4.5}$M$_\odot$ and galactic gas masses $>10^8$M$_\odot$
most of the ejected metals are retained by the galaxy. From
the bottom panel of Figure~\ref{fig:efa} we also see that a starburst
with a mass $>10^{5.5}$M$_\odot$ will eject more than 50\%\  of its
wind into the IGM.
\begin{figure}
\centering
\includegraphics[width=\columnwidth]{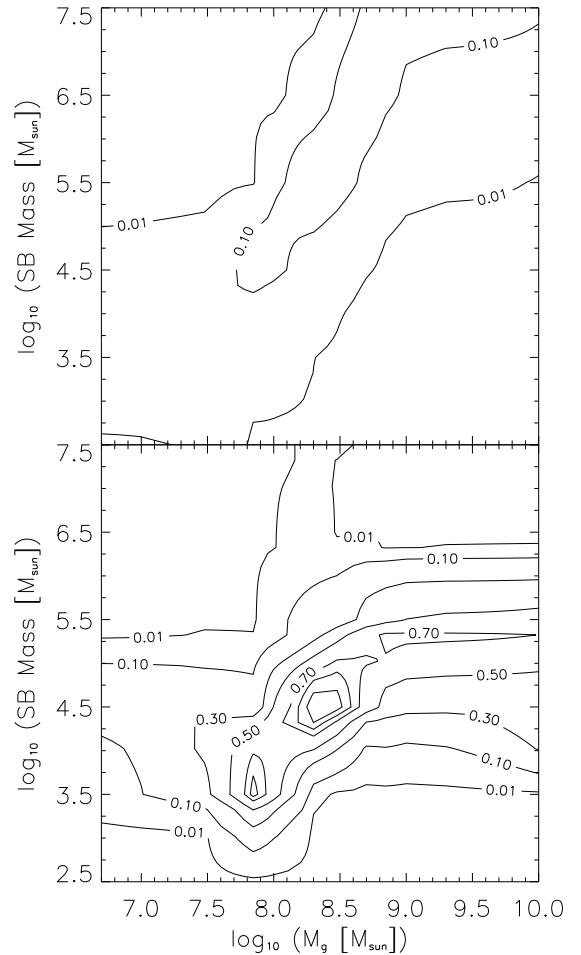}
\caption[]{Fractions of bound gas ($\chi_b$, top) and bound metallic gas
($\chi_{b,z})$ that reach an altitude of at least 3$H$, for the adiabatic
models.}
\label{fig:reta}
\end{figure}

Figure \ref{fig:efr} shows the $\xi_m$ (mass ejection) and
$\xi_z$ (metal ejection) efficiencies as a function of galactic gas
mass and starburst mass computed from the radiative models. Both the
$\xi_m$ and $\xi_z$ contour maps (top and bottom panels of figure
\ref{fig:efr}) are qualitatively similar to the ones computed
from the adiabatic models (see figure \ref{fig:efa}). Even though
the detailed shapes of the contours vary, the main difference between
the radiative and adiabatic cases is the following: in order to obtain similar
ejection efficiencies, one needs a starburst mass larger by a factor of $\sim
2$ in the radiative case.

\subsection{Material returned to the galactic disk}\label{sec:ret}
 
\begin{figure} 
\centering
\includegraphics[width=\columnwidth]{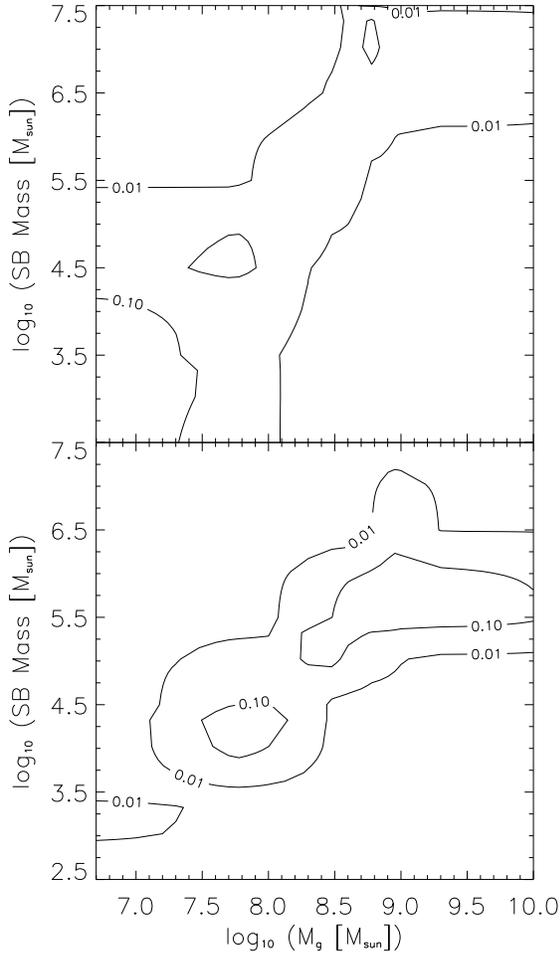}
\caption[]{Same as Figure~4, but for the radiative models.}
\label{fig:retr}
\end{figure}
Figure \ref{fig:reta} shows the efficiencies of gas 
($\chi_b$, top panel) and metal mass
($\chi_{b,z}$, bottom panel) that reaches at least 3 scale heights ($H$)
above the disk but remains gravitationally bound, computed from the
adiabatic models. Both panels show two low efficiency
regions divided by a high efficiency ridge (which runs diagonally from low to
high SB and galactic gas masses). In the high galactic mass/low SB mass
region, the SB ejection mostly remains close to the galactic plane,
and does not reach the required $3H$ height above the disk plane (in
order to be counted in the $\chi_b$ and $\chi_{b,z}$ efficiencies).
In the low galactic mass/high SB mass region, the SB ejection mostly
escapes from the galaxy. From the $\chi_{b,z}$ distribution (bottom
panel), we see that models with $M_g\sim 10^8\to 10^9$M$_\odot$ and
$M_{SB}\sim 10^3\to 10^5$M$_\odot$ are able to redistribute more
than $\sim 50$\%\ of the ejected metals over the galactic ISM.

Figure \ref{fig:retr} shows the $\chi_b$ (top panel) and $\chi_{b,z}$
(bottom panel) efficiencies computed from the radiative models. The
efficiencies show the same qualitative behaviour (with two low
efficiency regions separated by a high efficiency ridge) as the ones
calculated from the adiabatic models. However, the maximum values of
the efficiencies are lower. In the adiabatic case, the
maximum values of the gas mass  redistribution efficiency is of $\sim
20$\%\, while a maximum value of only $\sim 10$\%\ is found for the
radiative models. For the metal redistribution efficiency, values
above $\sim 70$\%\  are obtained for the adiabatic models, while
maximum values of only $\sim 20$\%\  are obtained from the radiative
models.

\begin{figure*} 
\centering
\includegraphics[width=14cm]{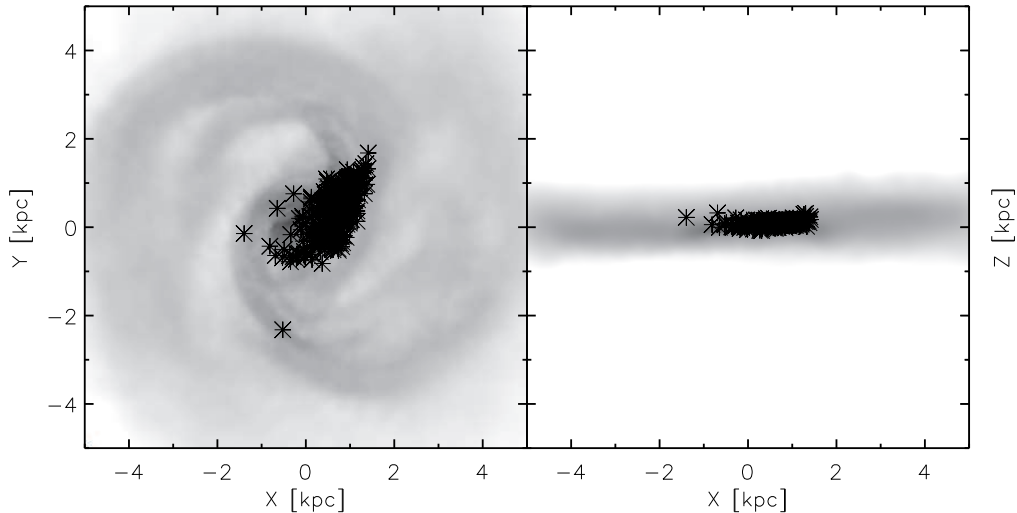}
\caption[]{Distribution of the SPH particles ejected by the
central starburst for model S7+G9 after 1Gyr. The inner region
of the domain is shown, and the location of the SPH particles
give the spatial distribution of the metals that have been
deposited back on the galactic disk. The left plot shows
the distribution of positions projected onto the galactic disk,
and the right plot shows the projection of the positions of the
SPH particles projected onto a plane which includes the rotation
axis of the galactic disk. The greyscale shows the distribution
of the mass density on these two planes.}
\label{fig:f6}
\end{figure*}
In Figure 6, we show the distribution of metals that have
fallen back onto the galactic disk from the starburst ejection
of model S7+G9, after a 1Gyr time evolution. The SPH particles
corresponding to the collapsed material form a distribution
with a height of $\sim 300$~pc (about the initial
scale height of the galactic disk), and with a radial extent
of $\sim 2$~kpc (see Figure 6), which is about twice the
size of the initial scale of length ($R_0$) of the G9 galaxy (see Table
1). This is an illustration of the fact that in all of our models
the infalling metals are redistributed over a large region of
the galactic plane.

\section{The effects of varying the radius of the injection
region and the resolution of the initial starburst}

In all of the models presented above, we have assumed that the
energy from the starburst is injected in a spherical region
of radius $R_c=150$~pc (which is a representative value of
the observed sizes of starburst regions, see \S 4.3). In this
section, we explore the effects on the ejection efficiencies
of varying the value of $R_c$.

We focus on three of the starburst+host galaxy combinations
which we have studied above (see Tables 1 and 2).
We choose models S5+G3 (a small galaxy with an intermediate
energy starburst, S10+G7 (an intermediate mass galaxy with
a strong starburst) and S11+G9 (a high mass galaxy with a strong
starburst). For these three models, we compute radiative simulations
in which the starburst energies are injected in spheres
of radii $R_c=50$ to 150~pc (in 20~pc steps). The number
of SPH particles within the spheres scales as the volume
of the initial starburst (45 particles for $R_c=50$~pc to
550 particles for $R_c=150$~pc).

In Figure 7, we show the mass ejection efficiency $\xi_m$ (see
Eq. \ref{efmass}) computed for models S5+G3, S10+G7
and S11+G9 as a function of the initial radius $R_c$ of the
starburst. We see that there is clear dependence of $\xi_m$
on $R_c$, with differences by factors of $\sim 1.7$ models
S11+G9 and S10+G7, and over $\sim 3.5$ for model
S5+G3.

We should note that the models with the smaller energy injection regions
(with $R_c=50$~pc) have smaller mass ejection efficiencies than
the models with larger energy injection regions (with $R_c=70$~pc,
see Figure 7). This is probably due to the fact that in the smaller
volumes (covering the denser, inner region of the galaxy) the
radiative cooling is stronger. In the models with even larger
energy injection regions (with $R_c\geq 90$~pc, see Figure 7), the
momentum per unit mass given to the galactic gas is lower, and
therefore less material succeeds to escape from the galactic potential.
From these results, we conclude that in our intermediate
and high mass galaxy models, variations of factors of
$\sim 2$ in the efficiencies are to be expected for starbursts
(with the same energies) with initial radii in the
$R_c=50$ to 150~pc range.

\begin{figure} 
\includegraphics[width=\columnwidth]{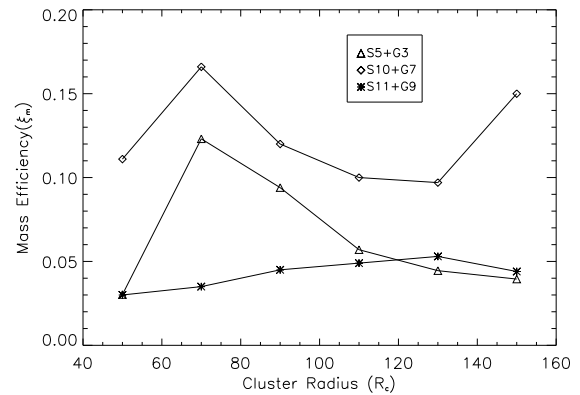}
\caption{The mass ejection efficiency $\xi_m$ computed for models S5+G3, S10+G7
and S11+G9 as a function of the initial radius $R_c$ of the starburst.}
\label{}
\end{figure}

We have also explored the effect of changing the total number
of SPH particles in the simulations. We again choose models
S5+G3, S10+G7 and S11+G9 (see above and Tables 1 and 2) as
representative of our model grid, keeping a constant,
$R_c=150$~pc initial radius for the starbursts, and
compute radiative simulations with  more than one order of magintud more of
gas, halo and disk particles (1500000, 2000000 and 1500000,
respectively) than the previously computed models (see \S 4).
In these new runs we have also changed the softening length to
$50~\mathrm{pc}$, for all the components.
In all the new models the mass ejection efficiency increases by a
factor of the order of $2$. For model S5+G3 this change is a factor of
$\sim 1.5$, while S10+G7 and S11+G9 suffer changes of factors of $\sim
2.21$ and $\sim 1.6$, respectively.

Therefore, we seem to obtain
at least a partial convergence for the values of the ejection
efficiencies in simulations with $\sim 10^6$ gas particles. However,
the computing time necessary for these simulations is quite large
($\sim 30$ times longer than the 30000 gas particle models), so that
our parameter study has been done at lower resolution.

\section{Discussion}

MacLow \& Ferrara (1999) computed radiative, axisymmetric simulations of
the ejection of gas from galaxies due to central starbursts. Their
models differ from ours in that the starburst energy is injected
as a continuous source term during the 50~Myr of their
simulations, while in our models the energy
of the starburst is injected instantaneously at the beginning of
the simulations.

If we convert their starburst mechanical luminosities into total
energies (by multiplying the luminosities by 50~Myr), and
using the relation between starburst energy and mass of Table~2,
we can find correspondences between their models and ours. We then find
that both the mass and metal ejection
efficiencies computed by MacLow \& Ferrara (1999) are
generally consistent with the ones that we have computed, with
typical differences of factors $\sim 3$. We judge this to be
a satisfying agreement given the differences between our models
and the ones of MacLow \&  Ferrara (1999).

Interestingly, the highest SB mechanical energy/lower galactic mass
model of MacLow \& Ferrara (1999) approximately corresponds to our
S5+G2 model (with M$_{SB}=3\times 10^4$~M$_\odot$ and
M$_g=3\times 10^7$~M$_\odot$). As can be seen in Figure 3,
this model lies within the transition from low to high efficiency
regimes. Therefore, the models of MacLow \& Ferrara (1999) mostly
missed the transition between low and high mass loss efficiencies
which we are describing in the present paper.

D'Ercole \& Brighenti (1999) computed a set of 5 models of
galaxies with $M_g<1.4\times 10^8$M$_\odot$, similar to our
galaxy models G1-G5, in which they inject energies in the range
of our S6-S8 starbursts. Their ``PEXT'' model (see Table 2 of
D'Ercole \& Brighenti 1999) has parameters close to our
S8+G4 model (see Tables 1 and 2). The PEXT model
has a $\xi_m=0.31$ mass ejection efficiency, which is
consistent with the $\xi_m=0.45$ efficiency obtained from
our S8+G5 model. D'Ercole \& Brighenti (1999) have studied
different possibilities (e. g., having galaxies with no rotation,
or having fixed dark matter mass distributions) which have not
been explored in our present work. 

Fragile et al. (2004) computed mass and metal ejection efficiencies
produced in galaxies with 10$^9$M$_\odot$ (similar to our galaxy
model G9) in which they inyect energies in the range of our model
S3-S7 starbursts. Their model~5 (see Table~1 of Fragile et al. 2004)
has parameters close to our S7+G9 model. Model~5 has a 
$\xi_m=0.016$ mass ejection efficiency and $\xi_Z=0.99$ metal ejection 
efficiency (see Table~2 of Fragile et al. 2004), which is consistent
with the $\xi_m=0.006$ and $\xi_Z=0.805$ efficiencies of our model
S7+G9. Fragile et al. (2004) have explored the metal ejection 
efficiencies for supernova events at different radial positions in
the galactic disk. They find that a substantial fraction of
the metals are retained in their off center starburst  models.  
These result shows that if we consider off-galactic center starbursts,
the mass and metal ejection efficiencies will be substantially lower
than the ones that we have computed in our present paper.

It is worth to mention that the strong winds produced by the
combination of individual stellar winds and SNe explosions not only
produce the GWs, but also have important consequences on the star
formation history of their host galaxy as well. They are the most efficient
way to redistribute metals to the ISM (and/or the IGM). This produces
a feedback mechanism that facilitates the gravitational collapse
of the ISM to form stars. The material with higher metallicity cools more
efficiently, and it therefore forms high density
condensations on shorter timescales than
metal poor gas  (Scannapiecco et al. 2005, Sutherland \&  Dopita 1993).
At the same time, the energy feedback from these strong winds can heat
and disrupt cold gas clouds, thus inhibiting star formation.

Observations and theory suggest (e.g. Larson 1974,
White \& Rees 1978, White \& Frenk 1991) that large outflows should be 
able to develop in small systems because of their shallower potential wells.
This is because, according to hierarchical galaxy formation scenarios, large 
systems are formed by the aggregation of smaller ones. Thus, the energy
feedback from strong winds should have a more important effect for systems
in the early stages of formation. 

Our models provide quantitative predictions of the amount of metals
which are ejected into the IGM and which are redistributed within the
host galaxy. These predictions could be used in a direct way for
computing models of the star formation history of the host galaxy,
and could also be included in models of the history of galaxy formation.

\section{Conclusions}

We computed 220 numerical simulations of GWs produced by
nuclear starburst energy injected inside a spherical region 
(of radius $R_c$) at the center of the disk galaxy. We used 
different total  masses of the galaxies G1-G10, combined with
different starburst  energies S1-S11, all adiabatic as well as
radiative (110 models each). We have performed adiabatic and radiative 
3D N-Body/SPH simulations of GW models using GADGET-2. the radiative
models have been calculated using a solar metallicity cooling
function. Finally, the adiabatic models were calculated in order
to obtain the maximum possible value of the mass
and metal ejection effciency in our models of galactic winds, and do
not necessarily correspond to real cases of starbursts in galaxies.

From the grid of  models we  obtained the mass and metal ejection 
efficiencies as a function of the mass of the gas in the galaxy and the total 
mass of the nuclear starburst.

In the same way, we computed the efficiencies of gas and metal mass
that reaches at least 3 $H$ above the disk but remains 
gravitationally bound for both adiabatic and radiative regimes. We present 
these efficiencies as a function of the mass of interstellar medium in
the galaxies versus total mass of the nuclear starburst.

We show that for compact starburst regions (e.g. inside the SSC) with masses 
larger than $10^6$ M$_\odot$ in the adiabatic case, the metal
injected by the nuclear starburst is expelled to the IGM (at least  
90\%\ of this gas). We show that in galaxies with gas 
masses between 10$^7$ to 10$^{10}$ M$_\odot$ and with a massive nuclear 
starbursts the available metals are freed from the gravitational potential 
of the galaxy. In the radiative case the critical SB mass at which total 
loss of metals occurs is around 2 or 3 times larger than the adiabiatic one.

In addition, we have calculated the efficiency of the metal redistribution 
in the disk of galaxies. We find that the metal redistribution is more
efficient for intermediate and low mass starburst masses ($<10^{4.5}$M$_\odot$)
and galactic gas masses $>10^8$M$_\odot$ (for both the adiabatic and radiative 
cases).

In this paper we presented  models in  which we 
do not consider many elements that are likely to be 
important in starburst galaxies, among these:
\begin{itemize}
\item clumpiness in the medium,
\item the generation of new stars,
\item a different fraction of gas in the galaxy disk,
\item the effect of galactic mass loading (see Fujita et al. 2009).
\end{itemize}

Therefore, there is a large range of possibilities for future studies
including differential and/or selective galactic winds (also, Pilyugin 1993 and
Recchi et al. 2001, 2002).

Finally, in our paper we have presented a limited resolution study (in
section 6). Our evaluation of the dependence of the computed
efficiencies on resolution is limited to a range of $\sim$3 in
resolution. This range is of course not very impressive, and
questions remain as to the results that would be obtained at
considerably higher resolutions (presently not possible, at least
for calculating relatively extended grids of models). When such
higher resolution models do become available, one will probably
find different numerical values for the efficiencies, but one would
expect that the transition between low and high ejection
efficiency systems (found in the present work) will be preserved.

\section*{acknowledgements}
We thank the anonymous referee for very relevant comments that resulted in a 
substantial revision of the original version of this paper.
We thank Volker Springel (from the Univ. of Heidelberg) for helpful
suggestions which we have been included in this paper. We acknowledge support 
from the CONACyT grants 61547, 101356 and 101975.

\end{document}